\documentclass[conference]{IEEEtran}
\IEEEoverridecommandlockouts
\usepackage{cite}
\usepackage{amsmath,amssymb,amsfonts}
\usepackage[ruled,norelsize,vlined,linesnumbered]{algorithm2e}

\usepackage{multirow}
\usepackage{graphicx}
\usepackage{textcomp}
\usepackage{xcolor}

\usepackage{url}
\usepackage{cite}
\usepackage{hyperref}
\usepackage{array}
\usepackage{textcomp}
\usepackage{stfloats}
\usepackage{verbatim}
\usepackage{subfigure}
\usepackage{pifont}
\usepackage{xcolor}
\usepackage{listings}

\usepackage{multicol}

\def\BibTeX{{\rm B\kern-.05em{\sc i\kern-.025em b}\kern-.08em
    T\kern-.1667em\lower.7ex\hbox{E}\kern-.125emX}}
\begin{document}
	
\title{Graphitron: A Domain Specific Language for FPGA-based Graph Processing Accelerator Generation\\
\vspace{-0.7em}
}

%\title{An Open Source Mixed-Precision Neural Network Accelerator Design Framework for FPGAs\\
%%\title{DeepBurning-MixQ: An Open Source Mixed-Precision Neural Network Accelerator Design Framework for FPGAs\\
%%\vspace{-0.7em}
%}

\makeatletter
\newcommand{\linebreakand}{%
\end{@IEEEauthorhalign}
\hfill\mbox{}\par
\mbox{}\hfill\begin{@IEEEauthorhalign}
}
\makeatother

%---------------------combined author format------------------
\author{
	\IEEEauthorblockN{Xinmiao Zhang$^{1,2}$, Zheng Feng$^{1,2}$, Shengwen Liang$^{1,2}$, Xinyu Chen$^{3}$, Cheng Liu$^{1,2}$\IEEEauthorrefmark{1}\thanks{\IEEEauthorrefmark{1} Corresponding author.}, Huawei Li$^{1,2}$, Xiaowei Li$^{1,2}$}
	\IEEEauthorblockA{
		$^{1}$SKLP, Institute of Computing Technology, Chinese Academy of Sciences, Beijing, China
	}
	\IEEEauthorblockA{
		$^{2}$Dept. of Computer Science, University of Chinese Academy of Sciences, Beijing, China
	}
 \IEEEauthorblockA{
		$^{3}$Microelectronics Thrust, Hong Kong University of Science and Technology (Guangzhou), Guangzhou, China
	}
	%	\IEEEauthorblockA{1120192664@bit.edu.cn, \{huanghaitong21s, liucheng, liguoyu21s,\\ wangying2009, lihuawei, lxw\}@ict.ac.cn, byang@hrbust.edu.cn}
	\vspace{-2.8em}

 \thanks{This work is supported by the National Key R\&D Program of China under Grant (2022YFB4500405), and the National Natural
Science Foundation of China under Grant 62174162.}
}

\maketitle
%\IEEEpeerreviewmaketitle

\begin{abstract}
FPGA-based graph processing accelerators, enabling extensive customization, have demonstrated significant energy efficiency over general computing engines like CPUs and GPUs. Nonetheless, customizing accelerators to diverse graph processing algorithms with distinct computational patterns remains challenging and error-prone for high-level application users. To this end, template-based approaches following established graph processing frameworks have been developed to automate the graph processing accelerator generation. Although these frameworks significantly enhance the design productivity, the templates often result in closely coupled algorithms, programming models, and architectures, severely limiting the versatility of the targeted graph processing algorithms and their applicability to high-level users. Furthermore, the limitations of the frameworks are usually ambiguous due to the absence of a rigorous grammar definition. 

To overcome these challenges, we introduce Graphitron, a domain-specific language (DSL), which allows users to generate customized accelerators for a wide range of graph processing algorithms on FPGAs without engaging with the complexities of low-level FPGA designs. Graphitron, by defining vertices and edges as primitive data types, naturally facilitates the description of graph algorithms using edge-centric or vertex-centric programming models. The Graphitron back-end employs a suite of hardware optimization techniques including pipelining, data shuffling, and memory access optimization that are independent with the specific algorithms, supporting the creation of versatile graph processing accelerators. Our experiments indicate that accelerators crafted using Graphitron achieve comparable performance to that generated with template-based design framework. Moreover, it exhibits exceptional flexibility in algorithm expression and significantly enhance accelerator design productivity.

\end{abstract}


\IEEEpeerreviewmaketitle

\section{Introduction}
% 图是一种重要的非结构化数据，许多人们面临的物理世界的数据问题，都可以利用图结构的来抽象表达，图计算已经被广泛的应用于大数据分析、网页搜索、社交网络、生物信息以及人工智能等重要的领域\cite{sahu2020ubiquity}。为了处理不断增长的大型图数据，主流的互联网巨头都投入并研发了自主的图计算引擎，如 Amazon 的图数据库 Neptune，Google的图计算系统 Pregel， Alibaba 的 MaxCompute，Oracle的PGX等。图计算系统已经成为核心的基础计算引擎之一。然而，目前主要的图计算系统基本上都是构建在通用处理器架构上也即CPU和GPU。尽管已经有了大量的研究工作，从高性能、可扩展性、可编程性等多个方面进行了深入的优化 ，这些基于软件的解决方案最终仍然受限于固定的通用硬件架构，通用处理器架构缺乏对于图计算本身特性的支持，图计算的能效难以进一步大幅度提升。

%对于图计算问题，通过体系结构的创新，设计图计算专用的体系结构是最有希望进一步大幅度提升图计算能效的方法。正如 Hennessy 和Patterson 在最近的技术报告中指出，通用的计算机体系结构发展逐步缓慢，领域专用的计算机体系结构迎来了巨大的发展机遇。以深度学习加速器为代表的领域专用加速器已经获得了巨大的成功，相比通用处理器，最新的深度学习加速器在性能和能效上都取得了数量级的提升。深度学习加速器的成功极大的激发了领域专用计算加速器的研究。图计算作为一个广泛使用的领域计算问题，也已经有了大量专用的图计算加速器研究工作，大量研究工作表明，专用图计算加速器也能大幅度提升图计算的性能和能效。

% FPGA（Field Programmable Gate Array）是一种硬件可重构的体系结构。它是在 PAL、GAL 等可编程器件的基础上进一步发展的产物。它是作为专用集成电路（ASIC）领域中的一种半定制电路而出现的，既解决了定制电路的不足，又克服了原有可编程器件门电路数有限的缺点 [11] 。FPGA 的可重构特性使得基于 FPGA 的图计算加速器可以通过对特定的图计算问题进行深度优化，从而有效的提高图计算加速器的效率。然而，FPGA 图计算加速的灵活定制也带来了新的设计挑战。为了对单个算法乃至具体的图进行深入的优化，就需要定制图加速器的设计。而硬件设计效率很低，开发周期长。专用硬件的加速器设计，通常以数月计算，而且需要大量的迭代优化。这样低效率的专用图计算加速器开发，难以满足图算法的多样化和广泛的使用需求。在这种情况下，敏捷的自动化图计算加速器综合就显得尤为重要。

Graphs have become increasingly important for representing real-world network data in critical fields, such as big data analytics, web search, social networks, bioinformatics, and artificial intelligence \cite{sahu2020ubiquity}.To address the growing demand for processing large-scale graph data, major Internet giants have invested in and developed their own graph processing engines, such as Amazon's Neptune\cite{neptune} graph database, Google's Pregel\cite{pregel} graph processing system, Alibaba's MaxCompute\cite{alibaba_maxcompute}, Oracle's PGX\cite{oracle_pgx}, and others. 
Graph processing systems have become one of the core foundational computing engines. % // 
%Although, there are extensive works exploring efficient graph processing systems on CPUs \cite{chen2019powerlyra} \cite{thinkinglkieavertex} \cite{kyrola2012graphchi} \cite{graphmat} and GPUs\cite{wang2023hytgraph}\cite{nvidia_nvgraph}\cite{gunrock}\cite{cusha}. 

%\cite{han2013turbograph}
Despite extensive research efforts\cite{nxgraph} \cite{ben2017groute} \cite{beamer2015gap}\cite{graphx} that have optimized these software-based solutions for high performance, scalability, and programmability, they are ultimately limited by poor hardware utilization due to the mismatch between architecture and irregularity of graph processing. For example, GPUs manage their threads in warp granularity, leading to several workload imbalance problems when traveling irregular and sparse graph structures\cite{gui2019survey}\cite{ceze2017democratizing}.

FPGAs, instead, allow customized hardware architecture for specific applications and have demonstrated high efficiency on graph processing problems~\cite{graphicionado}\cite{foregraph}\cite{ahn2015scalable}. 
Existing works usually leverage optimized code templates~\cite{ThunderGP,chen2022regraph} or hardware overlays~\cite{hu2021graphlily} that inherit memory access and pipeline optimizations tailored for graph processing domain to generate effective FPGA-based hardware accelerators for various graph algorithms. 
Although these graph processing frameworks have shown significant efficiency improvement on certain graph algorithms, their fixed optimizations or templates are often unsuitable for a wider range of graph algorithms. 
For example, an edge-centric graph processing framework could efficiently process all vertex active algorithms, such as PageRank, but struggles with traversal graph algorithms, such as Breadth First Search (BFS) and Single Source Shortest Path (SSSP), due to a large number of useless edges traversed. Furthermore, developers have to grasp the intricacies of the framework to express the graph task and generate high-performance hardware accelerators. 
These limitations have driven us to design a more flexible graph algorithm representation that necessitates a small learning curve and flexible design space exploration that composes hardware modules/optimizations adaptively for a wide range of graph algorithms on FPGAs.

% 在这项工作中，我们提出了一用于FPGA图计算加速器敏捷开发的领域专用语言 Graphitron。Graphitron使软件用户不必理解 FPGA 底层的复杂设计，仅通过特定的图操作函数和图操作符对图计算任务进行灵活描述，便可以为广泛的图处理算法生成定制化的FPGA高性能加速器。
% 这项工作的主要贡献可以概况如下：

In this paper, we introduce Graphitron, a domain-specific language for agile development of FPGA-based graph processing accelerators. 
Graphitron allows software users to flexibly describe graph processing tasks without engaging with the complexities of low-level FPGA designs. 
More importantly, the compiler of Graphitron can select and compose memory access optimizations or pipeline optimizations for the target graph algorithm during the compilation time. 
This capability expands the design space beyond that offered by existing approaches, allowing Graphitron to produce efficient hardware architectures for a wide array of graph algorithms.

Specifically, our work makes the following contributions:

\begin{itemize}
	\item We introduce Graphitron, a domain-specific language tailored for the rapid development of FPGA-based graph processing accelerators for the first time. Graphitron allows for the flexible description of graph processing algorithms and accommodate various programming models without engaging with the complexities of low-level FPGA designs.

	\item We have developed a compiler that automatically generates high-performance and end-to-end graph processing accelerators for FPGAs from graph tasks described in Graphitron. To augment the performance of these accelerators, the compiler integrates a set of hardware optimization techniques including pipelining, shuffling, and caching autonomously. Consequently, users can concentrate on algorithm design without the necessity to engage in hardware-specific optimizations. 
    
	\item According to our experiments on an Xilinx FPGA platform with five prevalent real-world graph  algorithms, Graphitron significantly advances both the computing efficiency and the accelerator design productivity. Notably, it achieves comparable performance to the leading template-based framework ThunderGP and manages to implement algorithms beyond the scope of ThunderGP.
\end{itemize}

\section{Related Work}
\subsection{FPGA-based Graph Processing Accelerator}
Graph has emerged as a critical data structure in a variety of big data applications, driving progresses in the hardware acceleration of graph processing \cite{nxgraph}\cite{graphx}\cite{chen2022regraph}\cite{su2024graflex}\cite{hu2021graphlily}\cite{hitgraph}\cite{graphmat}\cite{nvidia_nvgraph}\cite{cusha}\cite{gunrock}. 
%FPGA-based graph processing accelerators, which facilitate intensive algorithm-specific customization, are increasingly recognized for their superior performance and energy efficiency relative to general-purposed processors\cite{su2024graflex}\cite{chen2022regraph}\cite{hu2021graphlily}\cite{hitgraph} \cite{nxgraph} \cite{graphx} \cite{graphmat} \cite{nvidia_nvgraph} \cite{cusha} \cite{gunrock}. 
Among them, FPGA-based graph processing accelerators exhibit both superior performance and energy efficiency compared to general-purpose processors through algorithm-specific customization, thus attracting increasing research attention~\cite{foregraph}\cite{chen2022regraph}\cite{su2024graflex}\cite{hu2021graphlily}\cite{hitgraph}\cite{extrav}\cite{fpgp}\cite{attia2014cygraph}\cite{zhou2017accelerating}\cite{kapre2015custom}. 
%While graph processing has been notoriously difficult to optimize because of the low compute-to-I/O ratio and irregular data access patterns, extensive efforts have been undertaken to alleviate these issues.  
Since the performance of graph processing is severely constrained by its low compute-to-I/O ratio and irregular data accesses, extensive FPGA-based customization has been explored in previous works to mitigate these issues. 
FPGP\cite{fpgp} and ForeGraph\cite{foregraph} employed fine-grained partitioning with a tailored data placement to harness the high bandwidth and low latency of on-chip BRAMs for improving the efficiency of  random accesses in graph processing. 
% Zhou et al. \cite{zhou2016high} designed a graph data layout optimized to maximize the bandwidth of external memory accesses, thereby significantly increasing the throughput of graph processing. 
Cygraph~\cite{attia2014cygraph} customized the CSR format of graph and reconstructed the conventional BFS algorithm to maximize available memory bandwidth. 
%Cygraph et al. \cite{attia2014cygraph} and Kapre et al. \cite{kapre2015custom} leveraged multiple FPGAs for concurrent processing to achieve performance gains. 
Kapre et al. \cite{kapre2015custom} developed a graph-specific ISA based on abstracted sparse graph operations, and implemented specialized soft processors across multiple FPGAs. 
Zhou et al. \cite{zhou2017accelerating} explored the potential of CPU-FPGA heterogeneous architecture through mapping the vertex-centric processing and edge-centric processing onto the CPU side and the FPGA side, respectively, to accelerate graph processing algorithms with different processing paradigms. 
Lee et al. \cite{extrav} utilized FPGAs near storage to offload the generic graph processing operations, aiming to reduce the I/O overhead and optimize the efficiency of out-of-core graph processing. 
However, despite the high energy efficiency and performance, developing the FPGA-based graph processing accelerators remains a challenging task compared to implementing parallel graph processing engines with GPUs or multi-core CPUs.

\subsection{Design Productivity Challenges} 
Over the years, design productivity challenges has been alleviated from distinct angles through efforts from both industry \cite{xilinx_hls}~\cite{mentor_catapult_hls} and academia \cite{graphops}\cite{ftdl}\cite{wei2017automated}. % % 这里的引文做的不太好, 比如(1)是否起码应该引用一篇jason cong的论文? (2) 是否加起来的引用应该在5个以上? (3) 是否应该把industry和academia的efforts分类
%Many distinct approaches from both industry and academia have been proposed to address the design productivity challenges over the years. 
High-level synthesis (HLS)~\cite{xilinx_hls} encapsulates abstractions of hardware details into the grammar of high-level language to bridge the programming gap between high-level language and HDL, greatly lowering the barriers of FPGA design. 
%High-level synthesis (HLS) tools advancements, allowing software programs to be implemented on FPGAs with additional pragmas, have lowered barriers to FPGA design. 
%However, in practice, careful hand-crafted optimizations and a deep understanding of the transformation from application to hardware implementation are still required
However, implementing complex algorithms with intensive control logic and irregular memory accesses, such as graph processing, remains non-trivial when utilizing HLS~\cite{ThunderGP}. 
To this end, FPGA overlay \cite{ftdl} only exposes some specific functions or abstracts of the underlying configurable hardware accelerators to the high level language, allowing the application to fully exploit the high performance of FPGA conveniently~\cite{lee2016agile}.  
Especially, when deployed on deep learning processors or CGRAs, FPGA overlay has been demonstrated to significantly accelerate deep learning workloads or data flow graphs, respectively, with little programming efforts. 
%To this end, domain-specific overlay architectures \cite{ftdl} have been proposed to alleviate the design productivity and retain high performance at the same time \cite{lee2016agile}. 
%It has been demonstrated to be effective for applications such as deep learning and data flow graphs that can be deployed on configurable overlay architectures such as deep learning accelerators and CGRAs. 
%The basic idea is to decouple the problem to be compilation and configurable architecture which promises both competitive design flexibility and higher performance.
%However, FPGA overlay does not work well for graph processing, which can have distinct types of operations and computing patterns even though they may operate on the same graphs. 
However, FPGA overlay does not work well for graph processing, in which even processing the same graph, there can be distinct types of operations to and various computing patterns to the graph. % even though they may operate on the same graphs. 
%In this way, inspired by the various graph processing frameworks \cite{thinkinglkieavertex} \cite{shun2013ligra} \cite{powergraph} \cite{x-stream} that empower efficient parallel processing of different graph processing algorithms on the general purpose computing systems, template-based graph accelerator designs are proposed \cite{graphops} to provide both software-like programming flexibility and produce customized accelerators with higher performance. 
Inspired by various graph processing frameworks that enable efficient implementation of parallel graph processing algorithms on the general purpose computing systems, template-based graph accelerator designs are proposed~\cite{graphops}~\cite{hu2021graphlily} to provide software-like programming interface to customize high-performance graph accelerators. % ???加引用. Unlike the RTL-based templates,
To avoid programming with low-level HDL, ThunderGP \cite{ThunderGP} builds the templates with Xilinx HLS. 
Besides, ThunderGP enables the exploration to the design space, thus is more convenient to adapt to distinct graph processing algorithms. 
%However, ThunderGP is closely coupled with processing model like gather-apply-scatter (GAS) and programming models like edge-centric processing (ECP) which can be suboptimal for different graph processing algorithms. 
However, ThunderGP is closely coupled with gather-apply-scatter (GAS) graph processing model and edge-centric processing (ECP) pradigm, which can be sub-optimal for some graph processing algorithms. 
Additionally, users are required to modify the templates scattered across the different configuration files. 
Due to lack of rigorous definitions, this poses considerable challenges in formulation and verification of correctness. 

%Inspired by GraphIt\cite{zhang2018graphit}, which enables efficient parallel graph processing on multi-core CPUs and GPUs via a high-level domain-specific language (DSL), we propose Graphitron, aiming to agilely generate customized graph processing accelerators without low-level hardware architectural designs. 
GraphIt is a domain-specific language (DSL) for graph processing which generates efficient high-performance graph processing on multicore CPUs and GPUs \cite{zhang2018graphit}\cite{brahmakshatriya2021taming}. % 这里GPU还应该有一篇论文: taming the zoo...
This DSL flattened the learning curve of parallel graph processing by abstracting the most basic elements of graph processing into the language grammar, and generated high-performance implementation of graph processing by incorporating considerable optimizations. 
Inspired by it, we propose Graphitron, a DSL to agilely generate the customized FPGA-based graph accelerators without descriptions of any low-level hardware architectural designs. 
%Particularly, Graphitiron also investigate the typical graph processing acceleration optimization techniques such as data shuffling\cite{chen2019fly}, hub vertex caching, and pipelining integrated in the compiler backend to ensure the performance of the generated accelerators without compromising the flexibility of graph processing algorithm descriptions.
To ensure the performance of the generated accelerators, Graphitiron integrated typical FPGA-based graph acceleration optimizations into the backend of the compiler, such as data shuffling, hub vertex caching, and pipelining. % without compromising the flexibility of graph processing algorithm descriptions

\section{DSL for FPGA-based Graph Processing Accelerator Generation}

In this section, we will introduce the domain specific language, Graphitron, proposed for FPGA-based graph processing accelerator generation. The primary design principle is to allow users to focus on the description of graph processing algorithms without being aware of the underlying hardware architectural details. On the other hand, the compiler needs to address the hardware specific optimizations appropriately such that the resulting accelerators can still achieve competitive performance. We will start with the syntax definition, and then illustrate the overall compiler. Afterwards, we will introduce the major hardware optimization techniques that can be incorporated into the backend of the compiler. Finally, we will also explain the system integration work that is required to deploy the generated accelerators on a specific CPU-FPGA platform.

\begin{table*}[htbp]
\caption{Primitive APIs of generic graph processing algorithms.}
\label{algorithm language api}
\centering
\begin{tabular}{lll}
\hline
\textbf{Vertexset operators} & \textbf{Return Type} & \textbf{Description}                                              \\ \hline
size()                       & int                  & Returns the size of the vertexset.                                \\
init(func vp\_func)                 & none                 & Applies vp\_func(vertex) to each vertex to initialize vertex-related graph properties.            \\
process(func vp\_func)              & none                 & Applies vp\_func(vertex) to each vertex to process vertex-related graph operations.               \\\hline
\textbf{Edgeset operators}   & \textbf{}            & \textbf{}                                                         \\ \hline
size()                       & int                  & Returns the size of the edgeset.                                  \\
init(func ep\_func)                 & none                 & Applies ep\_func(edge) to each edge to initialize edge-related graph properties.                   \\
process(func ep\_func)                       & none & Applies ep\_func(edge) to each edge to process edge-related graph operations.      \\ \hline
\end{tabular}
\vspace{-1.0em}
\end{table*}

\subsection{Syntax Definition}
\begin{figure}[htbp]
	\centering	
\begin{lstlisting}
element Vertex end
element Edge end
const edges: edgeset{Edge}(Vertex, Vertex) = load(argv[1]);
const vertices: vertexset{Vertex} = edges.getVertices();
const old_level: vector{Vertex}(int);
const new_level: vector{Vertex}(int);
const tuple: vector{Vertex}(int);
const level: int = 1;
const activeVertex: vector{Vertex}(int);
func reset(v: Vertex)
    old_level[v] = -1;
    new_level[v] = -1;
end
func EdgeTraversal(src: Vertex, dst: Vertex)
    if (old_level[src] == level)
        tuple[dst] min= level+1;
    end
end
func VertexUpdate(v: Vertex)
    if (tuple[v]==(level+1)&(old_level[v]==-1))
        new_level[v] = tuple[v];
        activeVertex[0] = activeVertex[0]+1;
    end 
end
func VertexApply(v: Vertex)
    old_level[v] = new_level[v];
end
func main()
    vertices.`\textbf{init}`(reset);% Initialization
    old_level[1] = 1;
    new_level[1] = 1;
    var frontier_size: int = 1;
    while(frontier_size)
        edges.`\textbf{process}`(EdgeTraversal);
        vertices.`\textbf{process}`(VertexUpdate);
        vertices.`\textbf{process}`(VertexApply);
        frontier_size = activeVertex[0]; 
        activeVertex[0] = 0;
        level += 1;
    end
end
\end{lstlisting}
    \caption{Top-down BFS with ECP programming model}
    \vspace{-1.5em}
	\label{fig: bfs algorithm}
\end{figure}

Graphitron mainly employs graph data and properties to depict generic graph processing algorithms. The fundamental graph data types are vertex and edge, where edges can be either weighted or unweighted. An unweighted edge, containing a source and destination vertex, is represented by "(Vertex, Vertex)"; a weighted edge, which includes an additional weight parameter, is denoted as "(Vertex, Vertex, int)". Both vertices and edges can be aggregated to form a complete graph i.e. vertexsets and edgesets, which utilize curly braces "\{\}" to denote the data type of each element. Beyond these primitives, Graphitron includes a suite of basic graph processing algorithm description APIs, as listed in Table \ref{algorithm language api}, derived from the computational methodology in \cite{thinkinglkieavertex}. For instance, there are computing functions \textit{process()} defined on top of both vertices and edges, which can be applied for various graph processing models including vertex-centric programming model (VCP) and edge-centric programming model (ECP).

To illustrate the use of Graphitron, we take BFS as an example and present the code in Fig. \ref{fig: bfs algorithm}. BFS starts with an initialization operation with \textit{init} operator (Line 29) through \textit{reset()} function (Lines 10-13). It sets an initial value for each vertex's property in the vertex set. The \textit{process} operator defines the actions executed on each vertex or edge per iteration of graph processing, which are then executed on FPGAs for hardware acceleration. The compiler automatically forms the corresponding hardware kernels using Xilinx OpenCL. Multiple functions can apply to vertices and edges and execute in sequence, as demonstrated by Lines 34-36 in Fig. \ref{fig: bfs algorithm}. Function implementation details, such as \textit{VertexUpdate()}, \textit{VertexApply()}, and \textit{EdgeTraversal()}, are explicated in Lines 14-27. Graphitron accommodates both the ECP and VCP programming models simply by applying the corresponding functions on edges and vertices respectively. For instance, a VCP-based direction-oriented BFS example is shown in Fig. \ref{fig: DynamicSwitching}, which facilitates seamless transition between top-down and bottom-up BFS based on the frontier size (Line 10).

\begin{figure}[htbp]
    \centering
	% \centering	\includegraphics[width=1 \linewidth]{img/DynamicSwitching.png}

\begin{lstlisting}
func VertexTraversal(v: Vertex)
    if (old_level[v] == level)
        for ngh in v.getNeighbors()
            tuple[ngh] min= level + 1;
        end
    end
end
func main()
    ...
    if (frontier_size<0.05*vertices.`\textbf{size}`())
        vertices.`\textbf{process}`(VertexTraversal);
    else
        edges.`\textbf{process}`(EdgeTraversal);
    end
    ...
end
\end{lstlisting}
	\caption{Direction based BFS with VCP programming model}
    \vspace{-1.5em}
	\label{fig: DynamicSwitching}
\end{figure}

\subsection{Graphitron Compiler}\label{sec: compiler}
\begin{figure*}[tb]
	\centering	\includegraphics[width=0.85\linewidth]{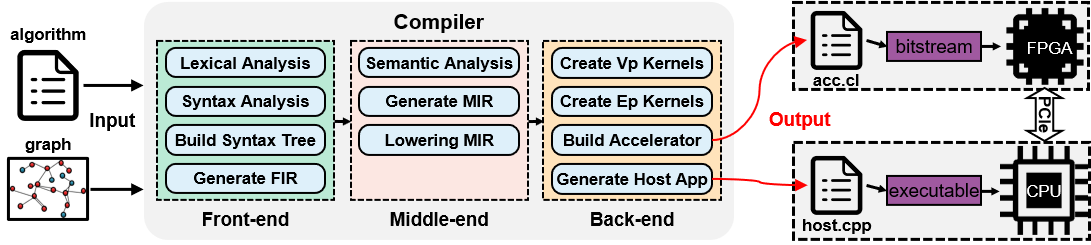}
	\caption{Overview of Graphitron Compiler.}
\vspace{-0.5em}
\label{fig:overview}
\end{figure*}

The compiler is crucial for supporting the syntax outlined in the preceding section. The architecture of Graphitron compiler is presented in Fig. \ref{fig:overview}. It accepts graph data and algorithm descriptions as inputs and fabricates high-performance FPGA accelerators via the classical compilation pipeline, including front-end, middle-end, and back-end stages. Particularly, rather than directly producing RTL code, it outputs Xilinx OpenCL code, thereby resolving low-level platform-specific details and simplifying the compilation process. The compiler's details are illustrated in the remainder of this section.

\subsubsection{Front-end}
%编译器的前端对算法描述进行词法分析和语法分析构建语法树和FIR上下文。我们以BFS算法中以边为中心编程模型的边遍历语句”#s1#edges.process(EdgeTraversal);“为例，如图5中(a)编译器首先对其进行了词法分析，将输入的字符流按照Token关键词扫描、识别为Token流，其中无法匹配关键词的字符串暂时定义为'IDENT'即标识符，如"edges"，这在语法识别时会进一步确定其身份，如图5(b)中在经过编译器的语法分析后，各个Token会按照语法规范组合为一个抽象的语法树，并按照语法赋予各个节点身份，如”edges“在语法树中充当着”ProcExpr“表达式的目标变量，其类型为”VarExpr“表达式，变量标识符名字为”edges“。示例中展示了一小段语法树，事实上我们的根节点Program下由多个”element“、”const“、”func“子节点构成，这些子节点又分别由多个语句、表达式、变量构成……总之，在编译器的前端，我们将用户的输入文件抽象为一种FIR上下文，并以语法树的形式存储。
% \begin{figure}[htbp]
% \begin{center}
%     \subfigure[TokenStream]{
    
%     \includegraphics[width=0.22\textwidth]{img/lexical_analysis.png}
%     }
%     \subfigure[Abstract Syntax Tree]{
%     \includegraphics[width=0.22\textwidth]{img/syntax_analysis.png}
%     }
% \caption{Lexical and Syntax Analysis}
% \vspace{-2.0em}
% \label{fig: lexical and syntax ananlysis}
% \end{center}
% \end{figure}

The front-end of the compiler performs lexical analysis and syntax parsing on the Graphitron code, subsequently constructs an Abstract Syntax Tree (AST), and generates the Front-End Intermediate Representation (FIR). Specifically, it scans the input file to generate a token stream based on recognized keywords. If the scanning process detects illegal expressions such as an unclosed string constant, it will throw an exception to signal the error. The front-end then parses the token stream and constructs FIRNodes of varying granularity, which comprise related tokens according to syntax rules. Finally, the front-end constructs the AST by assembling FIRNodes and exposes a pointer to the root node (\textit{Program}) within the FIR context, facilitating subsequent phases of the entire compilation.

\subsubsection{Middle-end}
%编译器的中间端对FIR执行语义分析以生成中间端中间表示(MIR)并降低MIR以使其更接近目标硬件代码的形式。编译器前端生成的抽象语法树(AST)是短视的，这意味着每个节点只能访问其直接子节点的信息。因此，中间端尝试从全局角度遍历语法树来执行语义分析，建立丰富的MIR上下文，如图形属性映射、符号表和函数映射，这使编译器能够更好地理解开发人员的意图。例如，考虑像“edges\textit{.process(update);”}这样的图操作语句。只有抽象语法树，编译器知道\textit{过程}操作符调用名为\textit{update}的图操作函数，但它不能访问\textit{更新}函数的参数、函数体或返回类型等细节。但是，通过MIR及其上下文，编译器可以很容易地检索这些信息。 从图操作内核、图属性和边集
%编译器的中端对FIR进行语义分析，同时获取用户自定义的硬件参数配置构建参数空间，根据参数空间和语义将FIR抽象为细节更丰富的低层次MIR。我们在前端中将用户的输入抽象为一个复杂、有序的语法树，但这个语法树它是”微观“的，即某一个节点只能获取它的子节点的信息，而无法从”宏观“的角度去理解整个图计算任务。在中端中，我们尝试从全局的角度重新从根节点遍历FIR语法树，在这个过程中我们根据节点信息建立元素类型映射、顶点和边属性映射、符号表以及函数映射等MIR上下文信息，最后以”宏观“的角度构建细节更为丰富的MIR，比如在MIR中我们对边集的加载、顶点和边属性的初始化、以及处理内核的构建等操作进行了lower，使得其不同于FIR，可以根据边集加载表达式的类别将边集存储为有权边集变量或无权边集变量，将{vector}定义的属性根据初始化表达式赋予不同的别名加入符号表中，根据{ProcExpr}表达式的目标类别创建边处理内核和顶点处理内核……这些操作更有利于上下文的变量调用，值得一提的是，我们在中端还会尝试获取用户定义的参数空间（如果配置了），并根据标签，为前文中所提到的边、顶点处理内核构建相应的参数空间，如图6所示，我们对{#s1#edges.process(EdgeTraversal);}配置了{program->configCacheBurstSize("s1", 7);}，因此该边处理内核的缓存突发读大小从默认的$2^{6}$被手动配置为了$2^{7}$，这使得我们具有对生成硬件内核手工调优的机会。
% \begin{figure}[htbp]
%     \centering
%     \centering	\includegraphics[width=1.0\linewidth]{img/ep_EdgeTraversal_kernel.png}
%     \caption{"EdgeTraversal" Edge Process Kernel}
%     \vspace{-0.5em}
%     \label{fig:ep_EdgeTraversal_kernel}
% \end{figure}

The middle-end of the compiler conducts semantic analysis on the FIR to create the Middle-End Intermediate Representation (MIR). It then lowers the MIR to more closely align with the form of the target hardware code. The AST produced by the front-end is limited as each node accesses only its direct children's information. To overcome this, the middle-end traverses the syntax tree from a global perspective to perform semantic analysis and establish enriched MIR contexts such as graph property mappings, symbol tables, and function mappings. These processes allow the compiler to interpret the developer's intent more comprehensively. For instance, from the statement "\textit{edges.process(update);}," the compiler identifies only the referral to the \textit{update} function via the \textit{process} operator. MIR, however, provides additional details like parameters, function body, or return type. 

Additionally, to bridge the software-hardware gap, the middle-end further adapts MIR for the target hardware, abstracting extraneous details outside of the developer's purview. For example, while software developers need not specify variable allocations to hardware kernels, such distinctions are vital for FPGA deployments. The middle-end employs a \textit{Property Detector} to determine the element and data types of graph properties within algorithms, aiding in the assessment of necessary memory allocation and designating where memory units are housed (either on the host or FPGA). It also automatically establishes memory unit IDs and indices for respective memory channels. These functions enable the FPGA to allocate memory and pointers for graph property transfers to the hardware kernel seamlessly, facilitating read and write operations on graph properties.

\subsubsection{Back-end}
The back-end of the compiler mainly generates the hardware accelerators based on the MIR context and deploys the accelerators on the target hardware platforms. Fig. \ref{fig:backend} reveals the overall framework required to bridge the gap between MIR and hardware platforms. It has a generic architecture to cover various graph processing algorithms, which is similar to a graph processing framework including all kinds of graph processing operations. These operations are then converted to hardware components implemented on top of Xilinx OpenCL. The components are categorized into four groups, each marked by a different color in Fig. \ref{fig:backend}. Blue represents user-specified hardware modules, reconstructed by the compiler according to middle-end MIR. Gray signifies static hardware modules that form essential components of the framework. Yellow denotes on-chip FPGA resources such as URAM, utilized for caching purposes. Finally, red signifies modules focused on optimizing memory access, leveraging various hardware techniques to enhance data access efficiency. Commonly shared by both edge and vertex processes, these operations aim to ensure efficient hardware implementation and versatile accelerator generation for diverse graph processing algorithms. 

\begin{figure}[htbp]
    \centering
    \centering	\includegraphics[width=0.8\linewidth]{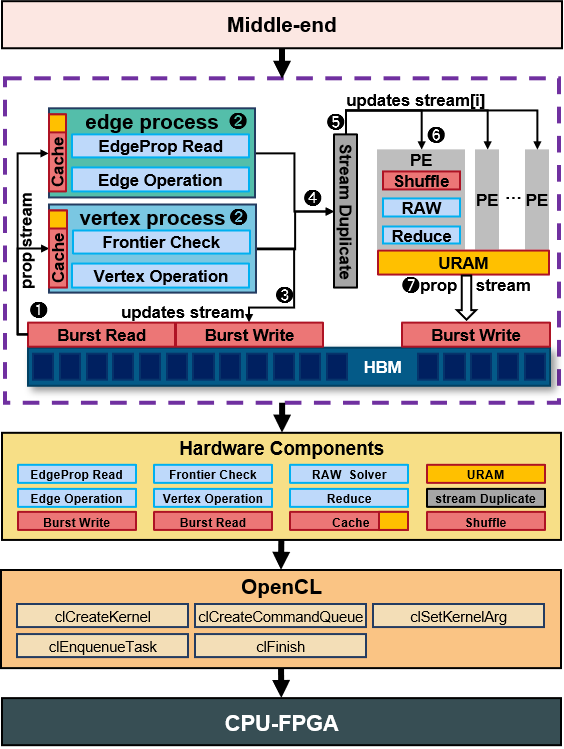}
    \caption{Back-end Framework}
    \vspace{-1.5em}
    \label{fig:backend}
\end{figure}

The overall architecture as shown in Fig. \ref{fig:backend} functions as follows. Graph data and properties are streamed into the \textit{Burst Read} module and then flow into the cache module for edge or vertex processing based on \textit{process} operator (step \ding{182}). In edge processing, graph data is read in the form of \textit{EdgeList}, and edge properties are selectively read through the \textit{EdgeProp Read} module based on whether there are edge weights. Then, \textit{Edge Operation} module outputs an updating stream containing vertex indices and update values based on the specific ECP graph operation (step \ding{183}). In vertex processing, graph data is read in the form of \textit{CSR}, and frontier are detected through the \textit{Frontier Check} module based on conditional statements of the graph operation. Then, the Vertex Operation module outputs updating streams based on the specific VCP graph operation (step \ding{183}). If the indices of the updating stream are sequential, the stream is written back to HBM using the \textit{Burst Write} module (step \ding{184}). Conversely, if the indices are unordered, the stream flows into \textit{Stream Duplicate} module (step \ding{185}). Then, the stream is duplicated and sent to different processing element (PE) units (step \ding{186}). In each PE unit, \textit{Shuffle} module reorders the update streams and filters out unnecessary data. Following the user defined reduction operations, data is written to the destination vertex properties in on-chip URAM cache without conflicts, facilitated by the \textit{RAW} resolver and \textit{Reduce} module (step \ding{187}). Finally, graph properties in URAM are written back to HBM using the \textit{Burst Write} module (step \ding{188}).

The major hardware optimization techniques such as pipelining and unrolling are mostly incorporated automatically, but they are critical to the performance of the generated accelerators and will be illustrated in Section \ref{sec: hardwareopt}. In addition, Graphitron provides an end-to-end graph processing acceleration system on a hybrid CPU-FPGA architecture, so there are also additional system integration requirements such as accelerator management, data transfer between CPU and FPGA, and graph preprocessing. This part is not part of the graph processing algorithm, so it is also made transparent to users and will be illustrated in \ref{sec: system}

%as the target platform and utilize Xilinx OpencL instead of RTL to simplify the complex platform adaption issues like accelerator drivers and communication facilities between host CPU and FPGAs. Eventually, we have the computing kernels deployed on FPGAs and some auxiliary code executed on host processors. It can be cumbersome for high-level users to determine the hardware and software partitioning tasks. Hence, we also need the back-end to make the decision implicitly and keep all the hardware details transparent to high-level users. This part will be illustrated in Section \ref{sec: system}. 

\subsection{Hardware Optimization Techniques} \label{sec: hardwareopt}
Since the graph processing algorithms to be accelerated are unknown at compilation time, algorithm specific optimizations will limit the description of Graphitron and undermine its flexibility. Therefore, we mainly explore generic hardware optimization techniques such as pipelining, unrolling, and memory optimizations  in the compiler back-end to enhance the performance of the generated accelerators without compromising the flexibility of Graphitron.

\subsubsection{Pipelining}
Pipelining is a prevalent hardware optimization technique that can significantly reduce memory accesses. However, it is limited to a single producer-consumer model. High-level code often contains write conflicts that impede pipelining. To overcome this challenge, the compiler resolves data dependencies through the introduction of temporary variables. Fig. \ref{fig:sssp ori} illustrates a common read-write conflict of \textit{SP} in the \textit{sssp} function, while Fig. \ref{fig:sssp decoupling} presents the corresponding rectified code.

\begin{figure}[htbp]
    \centering
	% \centering	\includegraphics[width=1 \linewidth]{img/DynamicSwitching.png}

\begin{lstlisting}
func sssp(src:Vertex,dst:Vertex,weight:int)
    SP[dst] min= (SP[src]+weight);
end
\end{lstlisting}
	\caption{The original sssp function}
    \vspace{-0.5em}
	\label{fig:sssp ori}
\end{figure}

\begin{figure}[htbp]
    \centering
	% \centering	\includegraphics[width=1 \linewidth]{img/DynamicSwitching.png}

\begin{lstlisting}
func sssp0(v:Vertex)
    tmp[v] = SP[v];
end
func sssp1(src:Vertex,dst:Vertex,weight:int)
    SP[dst] min= (tmp[src]+weight);
end
\end{lstlisting}
	\caption{The decoupled sssp function}
    \vspace{-0.5em}
	\label{fig:sssp decoupling}
\end{figure}

\subsubsection{Loop Unrolling}
Loop unrolling is another widely used hardware optimization technique that enhances spatial parallelism. Nonetheless, high-level code frequently encompasses write conflicts across loop iterations. The \textit{VertexUpdate} function within BFS, exemplified in Fig. \ref{fig: bfs algorithm}, serves as a typical case. Similar to the pipelining optimization, we introduce independent temporary variables for each iteration and implement a reduce operation to yield an equivalent output.

\subsubsection{Memory Access Optimizations}\label{sec: mem} 
We have implemented a suite of optimizations to improve the efficiency of both on-chip and off-chip memory access, which is critical to the performance of graph processing accelerators. A burst read/write module enhances the utilization of memory bandwidth for off-chip sequential access, as depicted in Fig \ref{fig:memopt}(a). For random access to off-chip memory, we have introduced an optimized cache module, shown in Fig \ref{fig:memopt}(b), that amalgamates repeated vertex accesses and supports prefetching to augment data reuse. For the on-chip memory access, as illustrated in Fig \ref{fig:memopt}(c), we utilize a data shuffling \cite{chen2019fly} module to distribute data across multiple on-chip memory banks, enabling conflict-free parallel computation and read/write operations.

\begin{figure}[tb]
    \centering
    \centering	\includegraphics[width=0.9\linewidth]{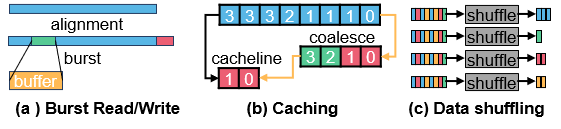}
    \caption{Memory access optimizations}
    \vspace{-0.5em}
    \label{fig:memopt}
\end{figure}   

% \begin{figure}[htbp]
% \begin{center}
%     \subfigure[Burst Read/Write]{
%     \includegraphics[width=0.282\linewidth]{img/burstread.png}
%     \label{fig:burst}
%     }
%     \subfigure[Caching]{
%     \includegraphics[width=0.309\linewidth]{img/caching.png}
%     }
%     \subfigure[Dada shuffling]{
%     \includegraphics[width=0.209\linewidth]{img/shuffling.png}
%     }
% \caption{Memory access optimizations}
% \vspace{-2.0em}
% \label{fig:memopt}
% \end{center}
% \end{figure}

% For sequential off-chip accesses, we have built burst read and write modules to fully utilize the high bandwidth of HBM on FPGA and improve the performance of the hardware accelerator. For example, when reading the source vertices of edges, we can align them on the host side and then read the source vertices of 16 edges at once in the kernel.For random off-chip accesses, we have implemented a caching module. As shown in Fig. for reading the source attributes, we can sort the edges by the source vertices in ascending order on the host side. This allows us to merge adjacent identical source vertices when reading the source attributes, reducing the total number of memory access requests. The caching module performs burst reads of multiple attributes after a cache miss for source vertex access.For on-chip access, we have designed a shuffle module to allocate destination vertices to different processing elements (PEs) based on the remainder of division, enabling parallel read and write computations across multiple channels.
\subsection{System Integration}\label{sec: system}

Graphitron offers an end-to-end system for graph processing acceleration on a hybrid CPU-FPGA architecture, necessitating considerable efforts to manage the accelerator, coordinate data transfer between CPU and FPGA, and preprocess graphs. Given the complexity of a pure hardware controller, we utilize the OpenCL framework for system integration. Specifically, we leverage Xilinx runtime (XRT) APIs such as \textit{clEnqueueMigrateMemObjects}, \textit{clSetKernelArg} and \textit{clEnqueueTask} to manage the data transfers and kernel execution on the CPU side. In addition, We also devised some implicit programming interfaces, for instances like graph loading, partitioning, and data migration between CPU and FPGA, that are demanded for system-wide integration yet remain obscured from developers. For example, before graph processing via FPGA kernels, graph data must be moved to the FPGA with a process transparent to developers. Additionally, to partition large-scale graph data, the compiler calculates the partition size $U$ based on URAM resources and organizes edges $(src, dst)$ into subgraphs with ascending src values within each subpartition. In summary, with the supporting interfaces, diverse hardware accelerators can be deployed conveniently, contributing to Graphitron's adaptability and flexibility.

\section{Evaluation}
%现在我们将从性能、额外开销以及设计效率等方面展开对Graphitron的评估，所有给出的结果都是基于真实的实现。
In this section, we compare the performance of Graphitron to that of the state-of-the-art template-based graph accelerator framework ThunderGP. Additionally, we analyze the effects of different performance optimization strategies and evaluate the design productivity of Graphitron.
\subsection{Experiment Setup}
\subsubsection{Hardware Platforms}
%我们将Graphitron部署在U280平台上，它由Xilinx Alveo U280 Data Center Accelerator Card和Intel Xeon E5-2680 V2 CPU组成。Alveo U280 板卡提供了2x16GB DDR4 DIMMs、8GB的HBM以及3个SLR，以此提供高达460 GB/s的带宽， 开发软件版本为Vitis 2019.2。
We implement the graph accelerators on the AMD Xilinx Alveo U280 FPGA board, which is equipped with 8 GiB of HBM2 capable of accessing 32 HBM pseudo channels via AXI3 interfaces. The synthesis, placement and routing, and simulation of the graph accelerators are performed using the AMD Xilinx Vitis 2019.2 suite, operating on a CentOS Linux 7 system. This host server is powered by an Intel Xeon E5-2680 V2 CPU and is equipped with 128 GiB of DDR3 memory.

%We deploy the graph accelerators on the AMD Xilinx Alveo U280 FPGA board, equipped with 8 GiB of HBM2 memory which could access 32 HBM pseudo channels through the AXI3 interfaces. We synthesize, place-and-route, and simulate the graph accelerators utilizing AMD Xilinx Vitis 2019.2 toolkits, which runs on CentOS Linux 7 hosted by a server with the Intel Xeon E5-2680 V2 CPU and 128 GiB DDR3 memory.  %The host OS is CentOS Linux 7. 
\subsubsection{Graph Datasets and Algorithms}
%我们选取了PageRank({PR})和Breadth-First Search({BFS})等经典图计算任务作为测试基准，使用了若干个不同大小的合成图和真实图作为加速器的测试数据集，具体数据集信息如表3所示。
We utilize both synthetic and real-world graph datasets as benchmarks. 
Table \ref{tab:datasets} lists the details of all the graph datasets in the evaluation. 
%Our evaluation is conducted  with a wide range of graph applications on real-world graphs.
\begin{table}[htbp]
\caption{The graph datasets.}
\label{tab:datasets}
\centering
\scalebox{0.95}{
\begin{tabular}{c||c|c|c|c}
\hline
Graph Dataset              & $|$V$|$    & $|$E$|$    & $Deg_{avg}$    & Graph Type    \\ \hline\hline
rmat-19-32(R19)~\cite{leskovec2010kronecker} & 524K    & 16.8M  & 32         & Sythetic  \\
HiggsTwitter(HT)~\cite{leskovec2014snap}           & 457K    & 14.9M  & 32.5       & Social  \\
wiki-topcats(TC)~\cite{leskovec2014snap}           & 1.8M    & 28.5M  & 15.9       & Web  \\
Amazon2003(AM)~\cite{leskovec2014snap}           & 403K    & 3.4M   & 8.4        & Social \\
pokec-relationships(PK)~\cite{leskovec2014snap}           & 1.6M    & 30.6M   & 18.8      & Social\\ \hline
\end{tabular}
}
\end{table}

%We attempt to use the same applications across different scenarios to study the impact of performance optimizations. 
We implement a variety of graph algorithms for evaluation, which composed of three classical graph processing algorithms including PageRank, Breadth-First Search (BFS) and Single Source Shortest Path (SSSP), as well as two state-of-the-art graph algorithms including Personalized PageRank (PPR) for network analysis and Calculate Graph Attention Weights (CGAW) for graph attention neural network. 
To avoid error-prone, PPR and CGWA are depicted in Algorithm~\ref{alg:personalized-pagerank} and Algorithm~\ref{alg:CGAW}, respectively. 
%To be more specific, personalized-PR (as depicted in Algorithm \ref{alg:personalized-pagerank}) is a variant of basic-PR, and CGAW (as depicted in Algorithm \ref{alg:CGAW}) is a crucial graph operation in graph neural networks.

%"CGAW"（calculate graph attention weights）
\begin{algorithm}[!h]
  %$PR_{old}[v] \gets \{0,score_{init},0,...,0\}$\;
  $PR_{old} \gets \{0,score_{init},0,...,0\}$\;
  %map[v] \gets \{0,1,0,...,0\}$\;
  $map \gets \{0,1,0,...,0\}$\;
  $m \gets 0.85$\;
  $\epsilon \gets 0.001$\;
  %\textcolor{red}{$m \gets 0.85$\;}
  \While{not all v have Converged}{
    \For{$s,d \in E$}{
    $contrib[d] \gets contrib[d] + \frac{PR_{old}[s]}{deg[s]}$\;
    }
    \For{$v \in V$}{
    $PR_{new}[v] \gets (1-m) \times map[v] + m \times contrib[v]$\; 
    % $PR_{new}[v] \gets (1-d) \times map[v] + d \times contrib[v]$\; % d是什么, 跟前面那个d重了? 是不是应该提前定义一下d
    \If{$|PR_{new}[v] - PR_{old}[v]| < \epsilon$}{
        $Converge(v)$\;
        %$Convergence$\;
    }
    }
    $Swap(PR_{new}, PR_{old})$\;
    }
  \caption{Personalized PageRank (PPR)}
  \label{alg:personalized-pagerank}
\end{algorithm}

\begin{algorithm}[!h]
  % \textcolor{red}{$w_{i,j} \gets exp(e_{ij})$ \;} % e_{ij}是什么? exp是什么?
    \For{$s,d \in E$}{
    $accum[s] \gets accum[s] + w_{s,d}$\;
    }
    \For{$s,d \in E$}{
    $attention_{s,d} \gets \frac{w_{s,d}}{accum[s]}$\;
    }
  \caption{Graph Attention Weights(CGAW)}
  \label{alg:CGAW}
\end{algorithm}

\subsubsection{The State-of-art Utilized in Evaluation}
ReGraph \cite{chen2022regraph}, GraFlex \cite{su2024graflex}, and ThunderGP \cite{ThunderGP} are considered as three leading FPGA-based solutions for graph acceleration. 
Among them, ReGraph combines the characteristics of algorithms and the architecture designs to accelerate both sparse and dense graph algorithms through a co-design approach. 
GraFlex mainly investigates scalable interconnection networks to support parallel high-performance graph processing among FPGA cores. 
At present, while Graphitron prioritizes generic memory access optimization in its back-end optimization, it lacks flexible network design interconnecting FPGA cores and comprehensive analysis to the characteristics of graph algorithm. 
Therefore, it is meaningless to compare Graphitron with ReGraph and GraFlex. 
Nevertheless, Graphitron aims to integrate these optimization techniques into its backend in the future. 

On the other hand, ThunderGP, as a framework for graph accelerator generation, investigates optimizing memory access from both on-chip and off-chip perspectives. 
In this way, we implement ThunderGP on Xilinx U280 as the state-of-the art design to evaluate the performance of Graphitron. Additionally, we also handcrafted HLS-based FPGA graph accelerator through software code without any HLS optimizations as the baseline. 
% Therefore, it is unfair to compare Graphitron with ReGraph and GraFlex, and we mainly compare with ThunderGP as a framework for graph processing accelerator generation that optimizes both the on-chip and off-chip memory accesses. 
% We implemented ThunderGP on Xilinx U280 as the state-of-the art design. Additionally, we also built graph processing accelerator using simple software code without any HLS optimizations and take it as the baseline. 

\subsection{Performance}
\begin{figure*}[htbp]
    \centering
    \centering	\includegraphics[width=0.9\linewidth]{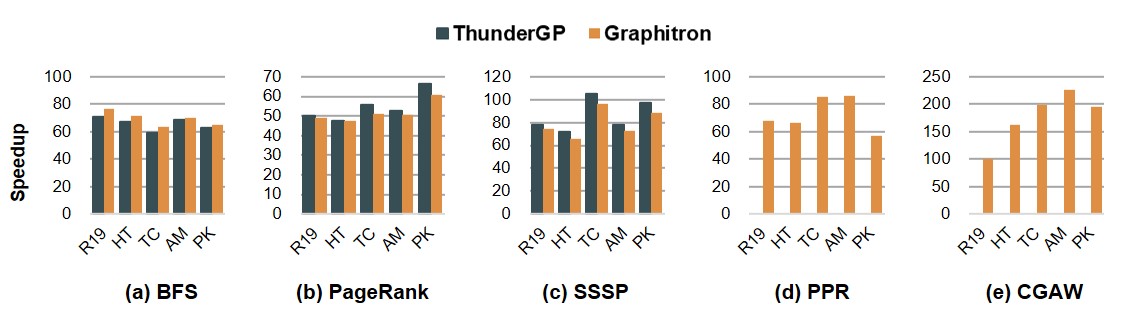}
    \caption{Speedups to the baseline of ThunderGP and Graphitron when running different graph algorithms.}
    \vspace{-1.5em}
    \label{fig:performance}
\end{figure*} % (1) 图里面baseline把B大写一下. (2) 如果baseline很低，已经看不见了，是不是可以考虑用对数坐标轴？(3) 纵坐标可以考虑改成speedup, normalized performance含义不明确 (4) 把thunderGP搞得颜色鲜明，graphitron不鲜明，为什么?  
\subsubsection{Comparisons with the State-of-arts}
As shown in Fig. \ref{fig:performance} which illustrates the speed-ups of Graphitron and ThunderGP to the baseline, both Graphitron and ThunderGP demonstrate notable performance acceleration. 
%In comparison to the baseline, Graphitron and ThunderGP achieve remarkable speedups of up to 86$\times$ and 226$\times$, respectively.
This is attributed to the integration of various FPGA hardware optimization strategies. 
Thanks to the flexibility of DSL which could switch between vertex-centric and edge-centric processing paradigms, Graphitron consistently surpasses ThunderGP on all datasets in BFS, with a maximum speedup of 1.09$\times$. 
%In BFS, Graphitron consistently surpasses ThunderGP on all datasets with a maximum speedup of 1.09x due to the flexibility of switching between vertex-cetric and edge-centric processing paradigms. 
%However, Graphitron's performance slightly lags behind ThunderGP in basic-PR and SSSP, , albeit consistently achieving over 90\% of ThunderGP's performance even in the most challenging scenario (SSSP on PK). 
However, the performance of Graphitron is lower than that of ThunderGP in PageRank and SSSP with consistently over 90\% of the speed-ups of ThunderGP.  
%This variance arises from Graphitron's end-to-end hardware implementation based on user input, contrasting ThunderGP's utilization of predefined functions within a fixed, high-performance hardware framework. 
This performance gap arises from that Graphitron implements end-to-end graph accelerator purely based on user-defined input, while ThunderGP utilizes encapsulated high-performance hardware interfaces with predefined functions. 
For instance, in PageRank, ThunderGP only supports returning the cumulative sum of differences between old and new PageRank values, but Graphitron enables developers to customize vertex properties to record these differences individually. 
This difference introduces optimization challenges in specific details, which degrades the performance slightly compared to ThunderGP. 
However, as for PPR and CGAW, Graphitron could effortlessly generate end-to-end graph processing accelerators based on its flexible descriptions and compilers, while ThunderGP which relies on the fixed interfaces in the template cannot accommodate these applications because of lacks support for additional graph properties required by PPR and the writing operations to the edge weights in CGAW. 
Graphitron accelerates the baseline by up to 86$\times$ and 226 $\times$ in PPR and CGAW, respectively. 
% \textcolor{red}{In comparison to the baseline, Graphitron and ThunderGP achieve remarkable speedups of up to 86x and 226x, respectively.}
%the end-to-end Graphitron's end-to-end hardware implementation based on user input, contrasting ThunderGP's utilization of predefined functions within a fixed, high-performance hardware framework. 
%For instance, ThunderGP's fixed template in basic-PR only supports returning the cumulative sum of differences between old and new PageRank values, but Graphitron enables developers to readily customize vertex properties to record these differences. 
%Furthermore, in personalized-PR and CGAW, ThunderGP cannot accommodate these applications due to its rigid hardware framework, which lacks support for additional graph properties required by personalized-PR and the writing functionality for edge weights in CGAW. 
%However, Graphitron effortlessly generates end-to-end graph processing accelerators based on flexible descriptions and compilers. 
% Overall, Graphitron achieves performance comparable to ThunderGP, with up to a 1.09x speedup in specific applications, and it can implement some algorithms that ThunderGP does not support.

\subsubsection{Performance of Different Memory Access Optimizations}
In this section, we evaluate the impact of different memory access optimizations implemented in the back-end of Graphitron. 
Specifically, we implemented different versions of Graphitron by incorporating distinct single memory access optimizations. 
%We denote the one integrating only burst optimization (Graphitron-withBurst), another solely caching (Graphitron-withCache), and a third exclusively shuffling (Graphitron-withShuffle). The normalized performance relative to the baseline (using BFS as a benchmark) is depicted in Figure \ref{fig:memoryperformance}.
In this way, we denote Graphitron which integrates only burst optimization, hot vertex caching and exclusive shuffling in the backend as Graphitron-withBurst, Graphitron-withCache, and Graphitron-withShuffle, respectively. 
The speedups to the baseline of different versions of Graphitron when running BFS are shown in the Figure \ref{fig:memoryperformance}.
\begin{figure}[htbp]
    \centering
    \centering	\includegraphics[width=0.85\linewidth]{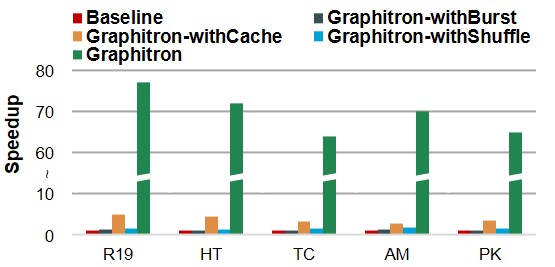}
    \caption{Speedups to the baseline with different memory access optimizations when running BFS.}
    \vspace{-0.5em}
    \label{fig:memoryperformance}
\end{figure} 
%Through comparative analysis, it becomes apparent that Graphitron versions implementing individual memory access optimizations demonstrate relatively modest performance improvements compared to the baseline. This is because relying solely on one strategy exposes other memory access patterns as overall performance bottlenecks. Specifically, Graphitron-withCache exhibits the highest performance enhancement, averaging a 3.8x speedup relative to the baseline. Subsequently, Graphitron-withShuffle achieves an average 1.4x speedup relative to the baseline, while Graphitron-withBurst shows the weakest improvement, with a maximum of only 1.1x. Noteworthy is the combination of all three optimizations in Graphitron, which mitigates various memory access performance bottlenecks, resulting in a substantial average performance acceleration of up to 69.5x relative to the baseline.
It demonstrates that incomplete Graphitron which integrates only individual memory access optimization exhibits only limited speed-ups to the baseline. 
This is because the performance of graph processing is bottlenecked by different irregular memory access patterns, thus optimizing only an individual memory access pattern still leaves other memory access patterns as the performance bottleneck. % ???
It can also be confirmed by that since the combination of all three optimizations mitigates the memory access issues from multiple perspectives, Graphitron leads to a substantial average performance acceleration of up to 69.5$\times$ to the baseline. 
Among all memory access optimizations, caching the hub vertices exhibits the highest performance enhancement with 3.8$\times$ average speed-up to the baseline, indicating that optimization which combines the characteristics of data with architectural designs can effectively enhance the performance of graph accelerator. 
%Besides, exclusive shuffling achieves an average 1.4$\times$ speedup to the baseline, while burst optimization shows the weakest improvement, with a maximum of only 1.1$\times$. 

% The experiment reveals the efficacy of Graphitron's memory access optimizations, highlighting that individual optimizations yield limited benefits, with significant performance gains realized only through their collective application.

\subsection{Design Productivity Comparison}
%图计算加速器的设计需要大量的底层电路设计和复杂的定制，往往只能依赖专业的硬件开发人员来进行，极大的制约了基于FPGA的图计算加速器的应用。图加速器的敏捷开发是图计算加速器设计的一个非常关键的目标，接下来我们从加速器敏捷开发的角度评估本文提出的基于领域专用语言图计算加速器设计方法，并对比了现有的基于模板的ThunderGP的图计算加速器设计方法。
%相对于以固定图计算框架为基础的FPGA图计算加速器敏捷开发方式，Graphitron使用户可以通过边和顶点的处理函数更自由的对算法进行描述，这使得用户不必在图计算框架的限制下对算法的表达做出妥协，而对于大多数无法按需修改的模板类加速器来说，用户只能对提供的框架模板进行简单修改，限制了对算法的表达方式。如ThunderGP需要用户理解Gather-Apply-Scatter的图计算模型，并在固定的函数和有限的参数下表达出自己的图算法需求，这对于一般用户来说是不友好的。
%另一方面，Graphitron在算法的表达上不受FPGA内核数量和类型的限制，用户可以在输入文件中调用边和顶点的操作函数按需描述，Graphitron编译器会自动根据描述生成对应的面向边和面向顶点的FPGA内核，并且还支持用户使用标签语言对不同的FPGA内核硬件参数进行手工调优，而ThunderGP中内核的数量和图计算流程都是固定的，用户只能根据模板填入参数，无法进行灵活的算法表达。
%此外，如图1和图2中Graphitron对于BFS算法的描述和参数空间配置都集中在同一个文件当中，而基于模板的工作中，如果需要对算法进行迭代更新往往需要修改多个文件。比如ThunderGP要实现与Graphitron同等的实现效果时，需要修改多达十多个文件，而且用户对其中一些文件的修改，需要对ThunderGP有深入研究并熟悉FPGA的硬件特性，这成为了敏捷开发的一个额外的门槛。
\subsubsection{Code Length Comparison}
Graphitron typically demands fewer lines of code in comparison to ThunderGP due to several factors. In ThunderGP, developers are required to abstract their graph algorithms into a push-based GAS model and then complete at least five application files based on templates. %In contrast, Graphitron allows for a succinct expression of algorithm description within a single file. 
In contrast, Graphitron allows expressing a graph algorithm within a single file. 
Furthermore, encapsulating all graph algorithms within the constraints of a graph framework and various parameters in the GAS model poses challenges, necessitating a profound understanding of graph algorithms and processing frameworks. 
Moreover, template-based configuration files of ThunderGP often entail comprehending underlying details of the FPGA and script construction methods for precise completion. 
Conversely, Graphitron empowers developers to precisely articulate their graph algorithm requirements using a concise and adaptable language, thus lowering the barrier to designing graph processing accelerators. 
For example, if a proficient user aims to describe a design in Graphitron by modifying templates of ThunderGP, it would entail adjustments to at least 14 files. 
%Conversely, if a proficient user aims to achieve a design akin to Graphitron by modifying ThunderGP's fixed templates, this process would entail adjustments to at least three sections of files totaling up to 14 files. 
Worse, altering certain files in ThunderGP, such as implementing a new hardware kernel, necessitates extensive manual debugging, leading to more complexity and difficulty in implementation.
\subsubsection{Algorithm Design Comparison}

% \begin{table}[htbp]
% \caption{Support for programming models.}
% \label{tab:programmingmodels}
% \centering
% \begin{tabular}{lccccccc}
% \hline
% System      &Vertex-centric  & Edge-centric & Hybrid    \\ \hline
% ThunderGP    &\ding{56} & \ding{52} & \ding{56} \\
% Graphitron &\ding{52} & \ding{52} & \ding{52} \\ \hline
% \end{tabular}
% \end{table}
Graphitron offers more agile algorithm design capabilities compared to ThunderGP. As shown in Table \ref{tab:algodesin}, Graphitron supports VCP, ECP and their hybrid models as graph processing programming models. 
In contrast, ThunderGP solely supports ECP, which traverses all edges irrespective of the frontier size. 
This could potentially lead to substantial bandwidth waste and computational overhead in algorithms with small frontiers, such as BFS.

\begin{table}[htbp]
\caption{Support of algorithms design.}
\label{tab:algodesin}
\centering
\begin{tabular}{lcccccc}
\hline
Systems       &vcp  & ecp & hybrid & weight  &kernels & properties  \\ \hline
ThunderGP    &\ding{55} & \ding{51} & \ding{55} &\ding{55} & \textbf{-}   &  \textbf{-} \\
Graphitron   &\ding{51} & \ding{51} & \ding{51} &\ding{51} & \ding{51} &  \ding{51}\\ \hline
\end{tabular}
\end{table}

Furthermore, in the template of ThunderGP, edge weights are assigned pseudo weights of random values, and it lacks support for loading weighted graphs. Besides, weights are treated as constants which cannot be assigned or modified by the accelerator. 
In contrast, Graphitron not only supports loading weighted graphs, but also empowers developers to dynamically modify edge weights during the runtime with algorithm description. 
This capability is crucial in algorithms related to graph neural networks, such as CGAW.

% Additionally, in ThunderGP's template framework, edge weights are assigned pseudo weights through random values, and it does not support loading weighted graphs. In contrast, Graphitron not only supports this operation but also allows developers to dynamically modify edge weights through algorithm descriptions. This capability may offer advantages in operations related to neural network training, such as adjusting weights based on node states. In ThunderGP, weights are treated as read-only attributes and cannot be assigned or modified.

%Moreover, ThunderGP's template-based design also imposes fixed constraints on both hardware kernels and graph properties, making it challenging to modify their numbers and types. While ThunderGP introduces optional vertex out-degree attributes to alleviate constraints on graph properties, it remains inadequate for algorithms like PPR requiring multiple graph attributes. In contrast, Graphitron generates synthesizable code end-to-end based on developer input, allowing flexibility in the number and parameters of hardware kernels. Graph properties can also be freely defined based on global variables (up to a maximum of 26 due to FPGA memory constraints), significantly enhancing develop freedom in accelerator design.
Moreover, template in ThunderGP also imposes fixed constraints on both hardware kernels and graph properties, making it challenging to modify their numbers and types.
While ThunderGP introduces an additional property of vertex out-degree to loose the constraints, it still remains inadequate for algorithms which requires multiple graph attributes, such as PPR. 
In contrast, Graphitron generates synthesizable code end-to-end based on developer input, allowing flexibility in the number and parameters of hardware kernels. Graph properties can also be freely defined based on global variables within the memory constraints of underlying FPGA, significantly enhancing freedom in accelerator design.

\subsubsection{Compilation Time Comparison}
%Concerning the hardware implementation time of the HLS-synthesized code, Graphitron requires additional 17 minutes compared to ThunderGP, accounting for 5.9\% of ThunderGP total duration. 
As shown in Figure~\ref{fig:timeoverhead}, the compilation time of Graphitron exceeds that of ThunderGP by 17 minutes, accounting for additional 5.9\% of the compilation time of ThunderGP. 
%This slight increase in time is attributed to Graphitron's support for custom hardware modules based on developers' flexible descriptions. For instance, introducing multiple graph properties adds additional memory pointers and memory access operations to the generated hardware modules, thereby slightly increasing complexity and resulting in additional software synthesis time overhead.
This slight increased overhead of Graphitron is mainly attributed to the supports for custom hardware modules based on developers' flexible design descriptions. 
For instance, introducing multiple graph properties adds memory pointers and memory access operations to the generated hardware modules, thereby slightly increasing complexity and resulting in additional software synthesis time overhead.
Besides, in contrast to ThunderGP, Graphitron incurs additional code generation time caused by compiler. 
However, the code generation time of Graphitron is only 0.115 seconds which takes only a very small fraction of the overall compilation time, which is negligible.
\begin{figure}[htbp]
    \centering
    \centering	\includegraphics[width=0.85\linewidth]{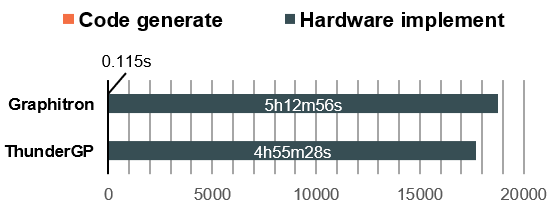}
    \caption{Complation time when implementing a BFS accelerator.}
    \vspace{-0.5em}
    \label{fig:timeoverhead}
\end{figure}

\section{Conclusion}
%\subsection{Conclusion}
In this work, we present Graphitron, a DSL for flexible graph processing accelerator generation on FPGAs without engaging with the complexities of low-level FPGA designs. Graphitron has vertices and edges defined as primitive data types and allows convenient graph processing algorithm description using with edge-centric programming models and vertex-centric programming models. In addition, Graphitron incorporates a suite of hardware optimization techniques such as pipelining, caching, and shuffling in the compiler back-end, thereby yielding autonomous accelerator optimization. According to our experiments on a set of representative graph processing algorithms, the accelerators generated with Graphitron achieve comparable to the state-of-the-art template-based graph processing accelerator design framework. Meanwhile, it also demonstrates exceptional algorithmic expressiveness and flexibility.

\newpage
\bibliographystyle{IEEEtran}
\bibliography{graphitron}

\end{document}